# Collective Transport of Lennard-Jones Particles through One-Dimensional Periodic Potentials

Jian-hui He, Jia-le Wen, Pei-rong Chen, Dong-qin Zheng and Wei-rong Zhong[1]

Siyuan laboratory, Guangzhou Key Laboratory of Vacuum Coating Technologies and New Energy Materials, Department of Physics, Jinan University, Guangzhou 510632, China

Abstract: Transport surrounding is full of all kinds of fields, like particle potential, external potential. Under these conditions, how elements work and how position and momentum redistribute in the diffusion? For enriching the Fick's law in ordinary, nonequilibrium statistics physics need to be used to disintegrate the complex process. This study attempts to discuss the particle transport in the one-dimensional channel under external potential fields. Two type of potentials – potential well and barrier do not change the potential in total – are built during the diffusion process. There are quite distinct phenomena because of different one-dimensional periodic potentials. We meticulously explore reasons about why external potential field impacts transport by the subsection and statistics method. Besides, one of the evidences of Maxwell velocity distribution is confirmed in assumption of local equilibrium. In addition, we have also investigated the influence of temperature and concentration to the collective diffusion coefficient, by a variety of external force, attaching thermodynamics analyze to these opposite phenomena. So simply is this model that the most valuable point may be an idea, which relating flux to sectional statistics of position and momentum could be referenced in similar transport problems.

## Introductions

There are three laws of transport in nature. The first one is Fourier's law, which describes how the heat diffuses in the media. The second is Ohm's law. We also know another kind of transport laws, Fick's first law, which successfully discovers the clear relationship between the particle flux and the concentration. If a transitory perturbation is given, a relaxation process would recover to equilibrium state in relaxation time. If a ceaseless force is given, the transport phenomenon is generated because of the ceaseless flux. To be different from other accurate analytical science, the Fick's law is about statistics, relating to the second law of thermodynamics. What drive particles? It is not particles themselves to push each other because each particle is independence, but only the probability![1] Further to speak, these knowledge and works of statistics might describe the social phenomena out of the box in untraditional physics application[2]. That is why diffusion is the field of statistics. A system tends to come back to equilibrium state when in nonequilibrium state, and balances the whole

---

[1] wrzhong@hotmail.com

distribution for more harmony.

Now the focus of statistics physics turns to nonequilibrium state. In biology, equilibrium equals death, nonequilibrium prevails in the creature, transport phenomenon is usual; the essence of chemistry process is nonquilibrium, there is no chemistry could happen in equilibrium. Nevertheless, so complex is solving problems about nonequilibrium state that we still depends on experiments, the essential element provided by the development of modern computer.

Let us disregard the fact that a particle is more or less affected by the force in the real world, and then so many researches based on Fick' laws without any force spring up firstly. Xin Liu et al obtain the Fick diffusion coefficients in the liquid mixtures of equilibrium state[3]. Z Chvoj has investigated diffusion transformations depending on the temperature[4]. Further on, the researches mixed with external force are becoming the hot points about the diffusion issue. Peter Prinsen and Theo Odijk elaborate the calculation of some parameters about collective diffusion coefficients of proteins considering the interacting of electrostatic and adhesive forces[5]. Alexander Tarasenko has discussed the collective diffusion in the one-dimensional homogeneous lattice[6].

However, the constrained motion is ubiquitous in a more intricate system or model. Ullrich Siems and Peter Nielaba study the diffusion and the transport of Brownian motion in two-dimensional microchannel combining periodic potentials and forces[7]. Given an issue like biological situation or soft condensed matter, both the potential barriers and the potential wells are the important factors operate inevitably so that the external force must be tangled in ionic channels of biology, nanopores and zeolites in materials science[8]. In the ion channel, ions would be accelerated or deaccelerated by the electric force, which caused by electric charge around the ion channel. In the two-dimensional tube, the shape of the tube is really the factor that affects the diffusion[9] because the particles would also be accelerated or deaccelerated in the progress direction by the reduction or addition in degree of freedom in the wide of tubes, let alone in the case of a deformable tube[10].

All in all, diffusion coupled with the multifarious and composite force is an interesting and necessary problem but still a challenge. The necessity is this can be found everywhere; the difficulty is the two fields, statistics physics is reflected in static state distribution and dynamics is reflected in particles transport, are hard to mingle together.

N. Koumakis et al. use the run and tumble model to describe kinetics' characters of particles while traversing energy barriers[11], but our work is a series of experiments completed by computer instead of such theoretical analysis. We perform particles transport behavior in external potentials by utilizing Monte-Carlo method and molecular dynamics. Briefly speaking, the particles – there are Lennard-Jones potentials between each particles – in the two places move freely as the impetus of concentration difference, what we call the Fick's law. There are also potential fields appended factitiously for the relationship that measures how the external potentials impact the diffusion coefficient.

This paper is organized as follows. Introduction in Sec. II is the model's schematic and some simulated elements, included external force field, L-J potential, Metropolis and Verlet algorithm, the Fick's law. In the Sec. III, there are relevant parameters what need to be simplified. Simulative graphs in Sec. IV acquire some interesting results of the diffusion coefficient gradual changings. Before the end of this paper, in Sec. V, it comes a conclusion and the applicability for further prospective.

# Schematic and model's principles

Especially in the subcellular level, the electric field should be considered in mass transfer because the road ions pass through the channel is not so smoothly on the account of the charge. Static electric fields[12] and alternating current fields[13] excited in the surroundings are the elements can change the diffusion coefficient generally. The diffusion is more complicated as the particles are affected by different forces in different positions. For this sample in biology, the transport phenomenon in the periodic tube[14] is another ambiguous and intricate collective diffusion problem. Meanwhile in the selectivity filter or protein channel,[15; 16; 17] the potential energy or the flies are particular the external force fields, and ions transport through that potential channel.

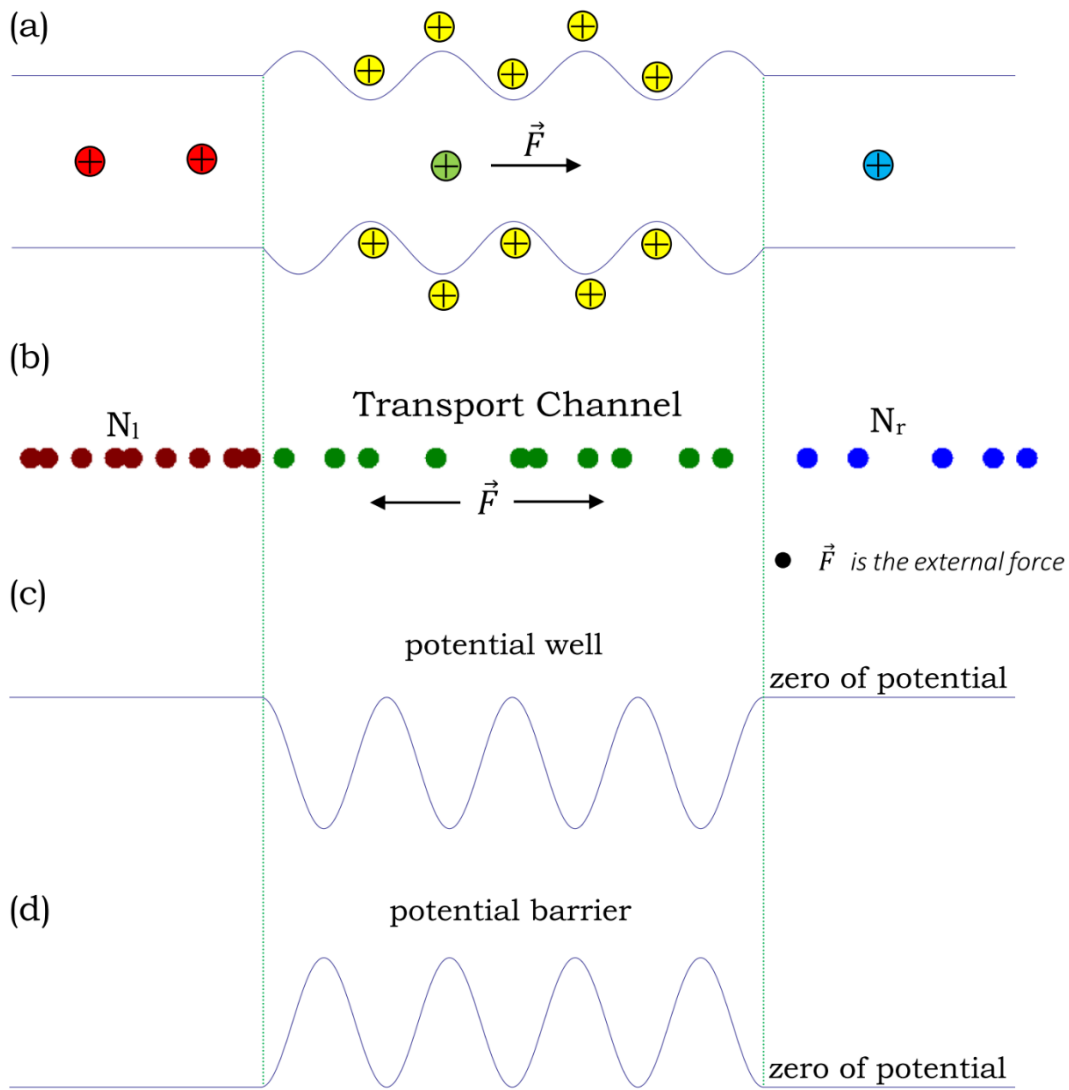

Fig 1: In the biological ion channel (usually like a corrugated tube), the ions are interacted inevitably by the charge around the channel (a). This system is consistent to a channel which a particle is passing through an external potential field (b). Usually, if the charge is electron then the external potential can be regarded as a potential well (c). On the contrary, if positive, the potential field is a potential barrier (d). The two ends of the channel are the particle reservoirs which maintain the concentration of the ions.

Particularly in the ionic corrugated tube, particles are stressed by the external force inevitably as we can see figure 1(a) displays the protomodel of ion channel observed usually in biological channel. The equivalent one-dimensional model shown in Fig.1(b) generalize this reality scene abstractly. In the schematic above, particles transport from the left side of high concentration to the right of low concentration, effected by the external alternating force, that equals particles transit the external fluctuate potential field up and down, then this channel is considered as the potential fields. The actual charge around the channel determines specific potentials to affect the collective diffusion. Provided that a positive ion is passing through the channel, if charges around channel are negative, as shown in Fig.1(c), the potential field is a well. Conversely, if positive, as shown in Fig.1(d), the field is a barrier. It is deserved to mention the potential in this model does not change the potential of particles in total, but it does change the diffusion coefficient, in other word, the diffusion ability.

Finite size effect is a normal phenomenon in many physical systems, such as thermal and electrical system[18; 19]. In this paper the one-dimensional model is used effectively. In the model, as shown in Fig.1, there are two particle reservoirs on the left and the right side separately. Between these particle reservoirs, there is a one-dimensional channel for the particles transporting between reservoirs to and fro. Therein, the fixed boundaries are rooted in the leftmost and the rightmost of the system. With the help of Monte-Carlo method, the number of particles in the two reservoirs can be maintained in a steady value. The particle-particle interaction is described as Lennard-Jones potential. The Langevin random heat baths[20] regulate the system's temperature and every particles' velocity, included reservoirs and channel, to become random variables distributed according to the Maxwellian distribution. The molecular dynamics and Verlet algorithm comes out with all kinetics parameters. The program sets 0.55fs as the length of every time step. As the feature, the particles transfer under the one-dimensional external potential, so the potential fields with multifarious parameters will be put into this channel and influence diffusion while particles move freely.

When the transport procedure is going, every particle in the channel is facing the external force from the force field. For representing circumstance generally, not only simple trigonometric functions of potentials like sin or cos functions but potential wells and barriers are introduced in the calculation process. To reflect two inverse conditions, the potential well means particles tend to drop into the potential field, and the barrier impedes particles from moving into the channel. The external potential is defined as $U_w$ and $U_b$ for the potential wells and barriers respectively. The mathematical form is

$$U_w(x) = -F_0 \frac{x_r-x_l}{2\pi n_p}\left[1 - \cos\frac{2\pi n_p(x-x_l)}{x_r-x_l}\right] = -\frac{1}{2}U_0\left[1 - \cos\frac{2\pi n_p(x-x_l)}{x_r-x_l}\right], \quad (1)$$

in which $F_0$ is the amplitude of the external force which be added factitiously; $n_p$ is the period of that force; $x_r$ and $x_l$ are the coordinates of the right and left side respectively of the channel; $U_0$ is the depth or height of the potential. It obtains the exact potential value in the exact place x. As the corresponding opposite field, the form of the potential barrier is the symmetry of the potential well is

$$U_b(x) = F_0 \frac{x_r-x_l}{2\pi n_p}\left[1 - \cos\frac{2\pi n_p(x-x_l)}{x_r-x_l}\right] = \frac{1}{2}U_0\left[1 - \cos\frac{2\pi n_p(x-x_l)}{x_r-x_l}\right] \quad (2)$$

Metropolis arithmetic, the most famous arithmetic in Monte-Carlo method[21; 22; 23], is the way to control the concentration in the particle reservoirs. Our method for maintaining the chemical potentials in the end regions uses stochastic particle creation and deletion trials every 50 time steps according to the prescription of grand canonical Monte Carlo (GCMC)[24]. In the one-dimensional reservoirs, the two ends are the immobilized boundaries and the amount of particles in reservoirs is set as N, then the system would probably create a particle if the amount is fewer than N. This probability is

$$P_{cr} = min\left[1, exp\left(-\frac{\Delta U}{kT}\right)\right], \quad (3)$$

in which k is the Boltzmann's constant, T is the particle reservoir's temperature, $\Delta U$ is an energy change after adding a particle.

Firstly, a random place of reservoirs would create a particle when amount less than N. If $\Delta U$ is under zero, the probability becomes 1, meaning the new born particle is acceptable and to give this particle a new coordinate and a new velocity which obey Maxwell distribution. If $\Delta U$ is above zero, it must create another random number P, which between 0 and 1. Only when $P_{cr} \geq P$, the new born particle is acceptable and given coordinate of there and a velocity according to the probability of $exp(-\Delta U/kT)$; When it does not, unacceptable, and delete this new particle.[25]

Reversely, the system would probably delete a particle if the amount of particles is more than N. The probability is $P_{dest} = min\left[1, exp\left(-\frac{\Delta U}{kT}\right)\right]$, and the operation is analogous to the situation which previous works have discussed.[26; 27]

Whenever managing realistic problems, it is inevitable to introduce interaction potential in different kinds of material. The Lennard-Jones potential[28] is applied in the free movement and diffusion in our simulation.

Molecular dynamics, that can be used to simulative diffusion[29], obtains dynamic data such as displacement, velocity, accelerated velocity, because the advantage of extensive applicability.[30] So modified way for Verlet algorithm[24] is utilized.

Moreover, the collective diffusion coefficient D, the important factor, bases on the meaningful diffusion law. This is Fick's law.

$$J = -D\frac{\partial c}{\partial L}, \quad (4)$$

In which D, the diffusion coefficient, represents the capacity of diffusion during the physics process and its unit is $\frac{m^2}{s}$; J is the mass flux of diffusion and unit is $\frac{mol}{m^2 \cdot s}$; c is the concentration and the unit is $\frac{mol}{m^3}$; L is the length of the channel.

## Definition of relevant parameters

The Lennard-Jones potential's parameters are regard as $\sigma_0$=0.258nm and $\varepsilon_0$=10.22K according to helium atom, m=1.67×10-27kg and k=1.38×10-23J/K. All of physical quantities in the simulative model need to be simplified: relative atomic mass $m^*$ is the magnification of one hydrogen atom; the length $L^* = L/\sigma_0$; the temperature $T^* = T/\varepsilon_0$; the concentration $c^* = c\sigma_0$; the potential parameters $\varepsilon^* = \frac{\varepsilon}{\varepsilon_0} = 1$ and $\sigma^* = \frac{\sigma}{\sigma_0} = 1$; the average velocity $\bar{v}^* = \bar{v}(m/\varepsilon_0 k)^{1/2}$; the potential $U_0^* = U_0 / \varepsilon_0 k$; the mass flux $J^* = J\sigma_0(m/\varepsilon_0 k)^{1/2}$; the collective diffusion coefficient $D^* = D(m/\varepsilon_0 k)^{1/2}/\sigma_0$.

## Results and discussion

For inquiring how different conditions work in the mass transfer, various transport procedures are under the control of different situations such as the average concentration and the differential concentration of reservoirs, the amplitude and the cycles of the external force, the length of the channel and the temperature of the system. Changing some conditions lead to similar diffusion trends no matter potential wells or barriers, but some changes lead to total disparate trends depend on the potential wells or barriers.

Here are some diagrams about the diffusion coefficient. In the simulation, the default number of steps is $8*10^8$, the temperature is 300K, the channel length is 100nm and the periodic number of the potential field is 6 if no postscript.

When the system stabilize the flux, it is the so called stationary state from then on. Nonequilibrium state is one kind of stationary state. Particles reach dynamical balance that means the system is in the minimal extreme value of Hamiltonian,

$$H_N = \sum_{i=1}^{N}\left[\frac{\vec{p}_i^2}{2m} + \phi(\vec{r}_i)\right] + \sum_{1 \leq i < j \leq N} u_{ij}(\vec{r}_i, \vec{r}_j) \quad (5)$$

where $H_N$ is the Hamiltonian of N particles system in the channel; $\frac{\vec{p}_i^2}{2m}$ and $\phi(\vec{r}_i)$ are the kinetic energy and potential energy respectively for the particle in the location $\vec{r}_i$ alone, and $u_{ij}(\vec{r}_i, \vec{r}_j)$ is the interactional energy between particles.

There are complete differences in such two potentials, and become larger and large when enhance the external field, but become no distinction in no external force since the absent of external field.

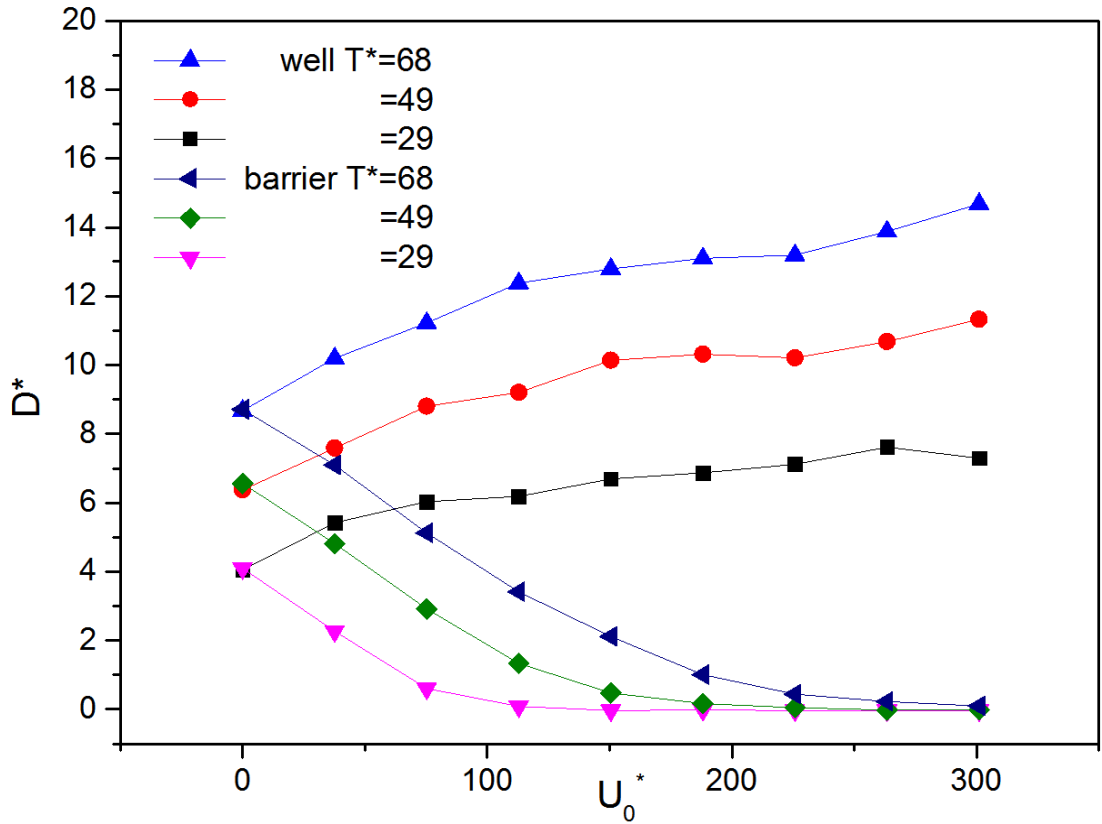

Fig 2: This is the dependent relation between the external potential field and the diffusion coefficient under the temperature of 300K, 500K and 700K. From the diagram, potential wells and barriers coincide when the potential is 0. Such two inverse conditions, one is good for particles motion and the diffusion coefficient, the other one hinders particles to transport so that the potential barrier decreases the diffusion coefficient.

In each N particles system of nonequilibrium state, a 2iN dimensional phase space describes that system where i is degree of freedom, N parameter of them are locations, N of them are momentum

(velocity, if in the same kind particle case). Acquiring the reason behind the picture why system in stationary state of different external potentials has different flux tendency, it is necessary to introduce a core equation: $J = \rho \cdot \bar{v}$, in which J is the flux, $\rho$ is the density, $\bar{v}$ is the average velocity of particles. Particle's flow makes the spatial distribution even; Particle's collision makes the velocity even. Because of nonequilibrium stationary state, $J, \rho, \bar{v}$ are constants depended on different environment for the coherent principle.

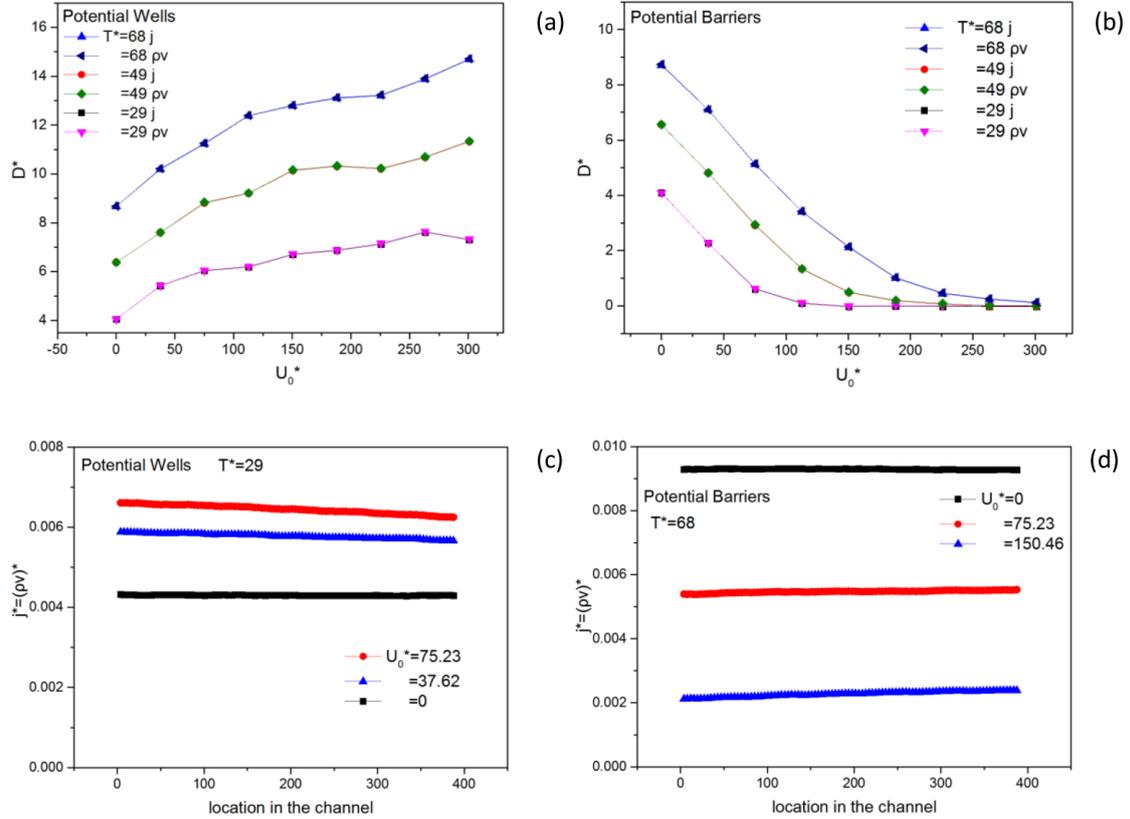

Fig 3: (a) We use the density and velocity of each small parts of the channel in the potential well. For comparing the flux—J, multiply density and velocity to obtain the other flux of each part, then get the average of a whole channel. The two flux obtained from different ways are exactly coincident. (b)That result also occurs in the potential barrier. This is the way we verify $J = \rho \cdot \bar{v}$ in the transport with external field. Besides, in the transport channel of potential well (c) and barrier (d), the product $\rho \bar{v}$ of density and velocity in the whole channel is still almost a constant.

In the assumption of local equilibrium, $\rho(\vec{r}, t)$ and $\langle \vec{v}(\vec{r}, t) \rangle$ are the functions to the variables of location and time. From pictures can see $\rho \bar{v}$ equals flux precisely. Diffusion process with external force is too abstract to analysis in theory, but this equation gives an idea to divide flux into density and average velocity. Now an analysis of particle distribution and velocity statistics could explain how external force impact diffusion.

Introducing particle distribution picture may be a visual, effective or better way to comprehend diffusion with potentials. In accordance with particle distribution below, the shape of that graphs can almost correspond completely to the mathematical function of potentials.

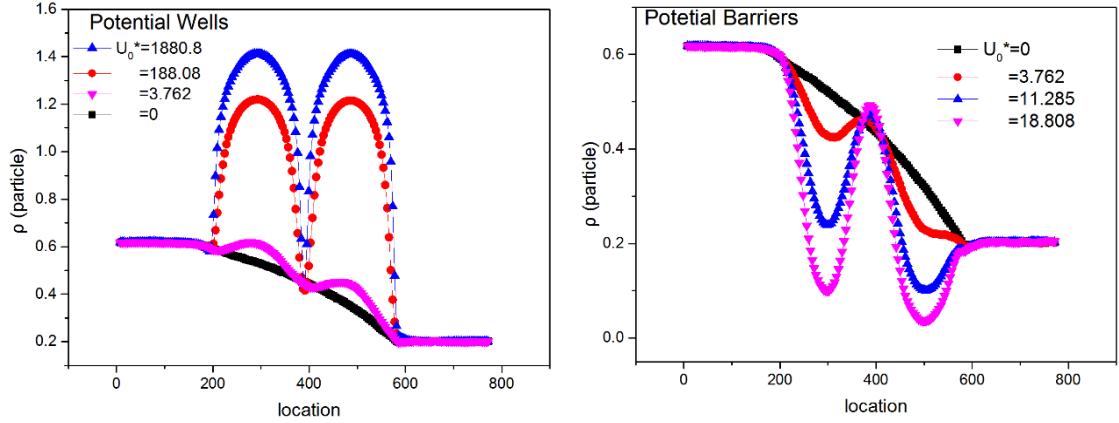

Fig 4: Here are pictures about the density distribution of particles in the 200nm long model. There are 200 pieces in total segmented from the model and 100 pieces of them segmented from the channel. The periodic number of the potential field is 2.

Two sides are the particle reservoirs of constant concentration. If comes no force, the distribution is very smooth from the high concentration to the low without any fluctuation. When the force is not so large, the distribution becomes the shape of the external field, but still press close to free diffusion; on the other side, when the force is considerable, the shape is dominated by the external field because of the probability distribution which we have introduced[31], and all peaks are on the horizontal line.

From the graphs of distribution above, the force amplitude is supposed to competes with differential concentration, based on the clue that the distribution approaches the potential shape when strengthen the force. On the other hand, the distributions are total different while potential is barrier or well. When the force is enhanced, the potential barrier arches more pointedly and the well sinks more deeply, so the barrier stores less particles and the well more. This assumption is the reason why the collective diffusion relies on the type of the potential field and its force amplitude.

If particles is propelled only by the concentration gradient with no external force, it is nonequilibrium condition completely. But concentration gradient can be ignored as detailed balance when the external field is large enough to the concentration gradient. Canonical ensemble distribution function is supposed to be dominant then problem can be managed by the equilibrium theory in the microscopic local scope. In such nonequilibrium stationary state and its constant distribution, particles which flow into a local position equals the same amount of flowing out particles; particle amount has a theoretical number and fluctuates around this value. On the equilibrium state, there is a function about the probability distribution to describe the population of identical particles

$$\rho(E) = \frac{1}{QN!} \exp\left(-\frac{-\Delta W(x)}{kT}\right) \quad (6)$$

$$Q = \frac{1}{N!} \int \exp\left(-\frac{-\Delta W(x)}{kT}\right) d\Omega \quad (7)$$

in which, k is the Boltzmann's constant, T is the medium temperature, N is the number of all particles, x is an arbitrary set of coordinates and W(x) is the minimum required reversible work changing the state[31]. Actually, the W(x) includes L-J potential to repel each particle mightily, and that is why to depend the simulation result.

For simplification as possible in the local scope. Particle velocity distribution is approaching Maxwell velocity distribution in equilibrium.[32]

$$f^{(0)}(\vec{r}, \vec{v}, t) = n \left(\frac{m}{2\pi kT}\right)^{\frac{i}{2}} \exp\left[\frac{-m}{2kT}(\vec{v}-<\vec{v}>)^2\right] \quad (8)$$

In which i is degree of freedom; m is the particle mass; T is temperature; k is the Boltzmann's constant; $\vec{r}$ is location; n is local density; $\vec{v}$ is velocity; t is time but stationary state is independent of time; $<\vec{v}>$ is the local average velocity. The average transport velocity is only several meters per second, and the local average velocity is just less than thirty meters per second. Both of them less than the usual particle velocity, which is why to omit the local average velocity and compare with the Maxwell velocity distribution directly.

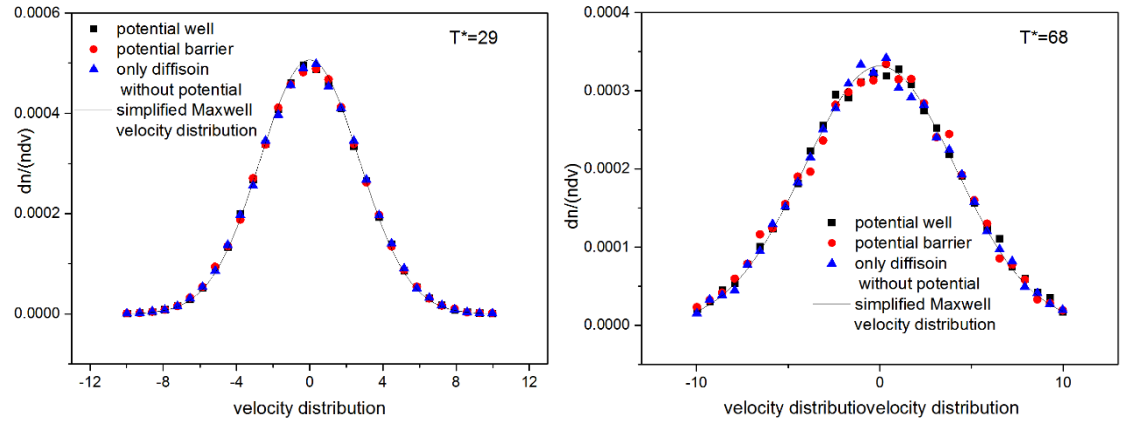

Fig 5: In the velocity distribution of potential well, potential barrier, and only diffusion without potential, these 300K and 700K temperature conditions of those are the same form to the Maxwell velocity distribution.

Now Maxwell velocity distribution which also for non-ideal gas, is confirmed exactly in the transport under the low drift velocity condition. The possibility of particle velocity accords with Maxwell velocity distribution no matter there are external potential, internal force and average drift velocity.

The entire average transport velocity is decided by temperature and concentration difference prevailingly. Next, the average velocity statistics of the entire transport comes.

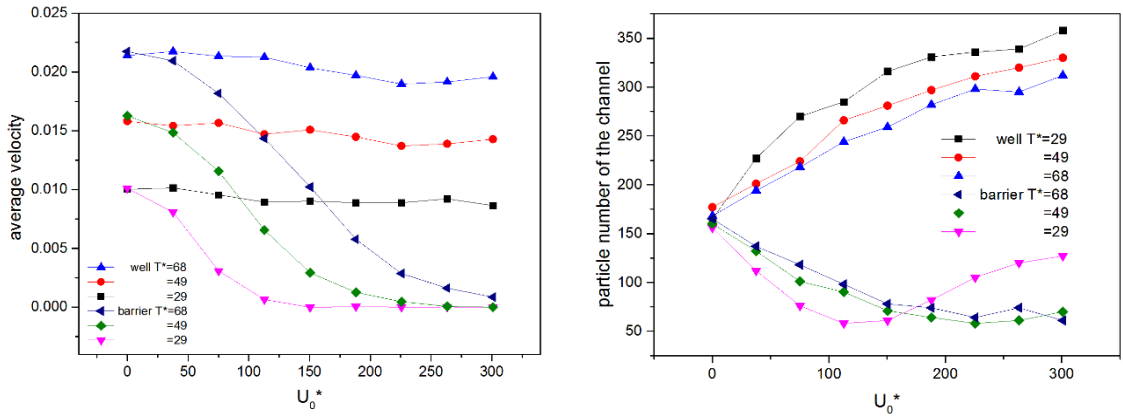

Fig 6: From the diagram of average, the higher temperature it is, the higher average velocity it has. Besides, the potential strength has little effect to the velocity in the potential well. In the particle number diagram, the changes of temperature and potential type are accordance with tendency in the Boltzmann distribution. Totally, density is the main factor for diffusion.

Then jump to the conclusion that once the external potential is introduced, according to that statistics mechanics, the higher energy level gets more particles, and the lower gets fewer, even though mutual L-J potential impedes to some extent. Because the potential well is the lower energy level than two reservoirs, there are more particles occupy wells, meaning the potential well is the surrounding made for mass transfer. In the other contrary surrounding, potential barriers are occupied by fewer particles because of its higher energy level, so that implies it is a hindrance to transport. Because of $J = \rho \cdot \bar{v}$, the larger of density comes with the larger of flux. On the other hand, for potential barriers, decreasing the external potential reduces both two variables – density and average velocity.

Some other works about other influence factors have been done. Even though some tendencies are similar, there are definite performances depend on the external force.

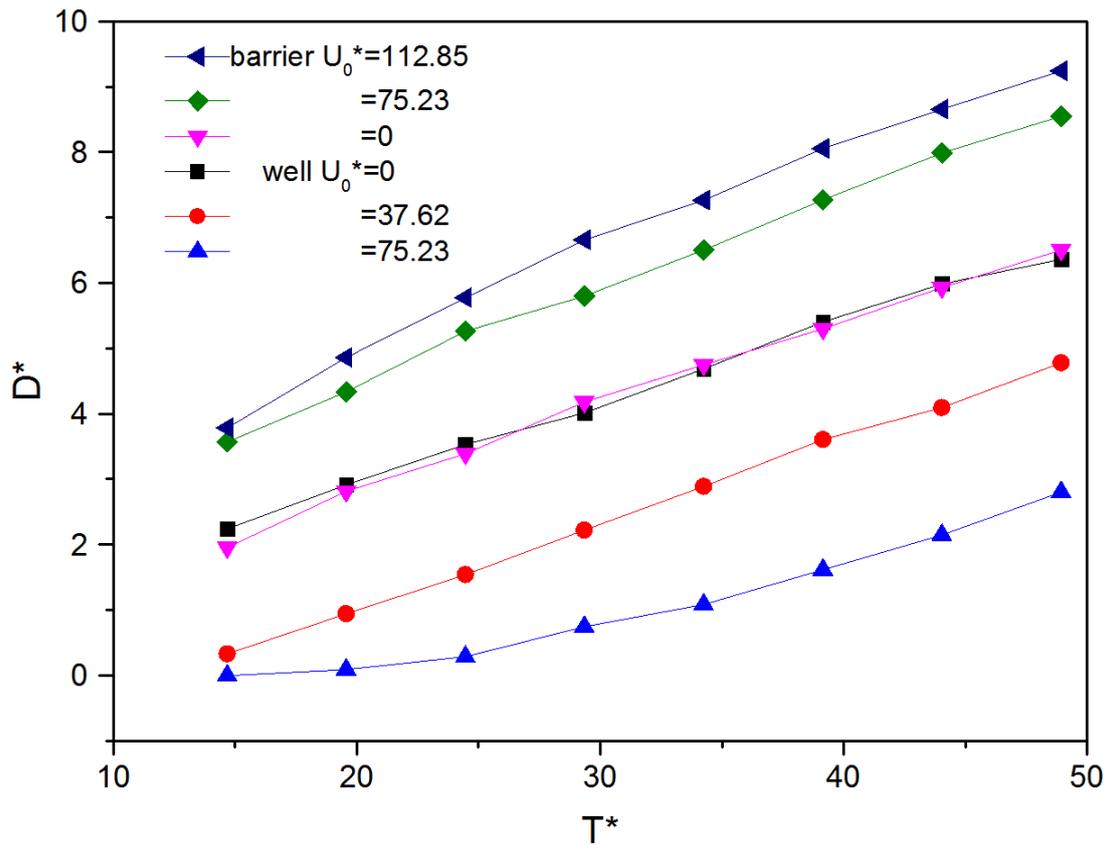

Fig 7: Temperature dependence of the collective diffusion coefficient for different potentials.

This picture attempts to illustrate the relationship between temperature and diffusion coefficient under different potentials. The range of the temperature is from 150K to 500K. The diffusion coefficient is increased in direct proportion to the temperature no matter wells or barriers. To potential wells, the diffusion coefficient is increased with the rise of the force amplitude; to barriers, the coefficient is decreased with the rise of the force amplitude. When amplitude becomes 0, two lines coincide.

It is not a difficult job to understand why the warming temperature is favorable for diffusion from the particle activity and rate. Otherwise, Einstein has gave a formula about the dissipation and the position fluctuate in 1905. $\zeta D = kT$, where $\zeta$ is the frictional coefficient. It clearly denotes the D is linear with the temperature.

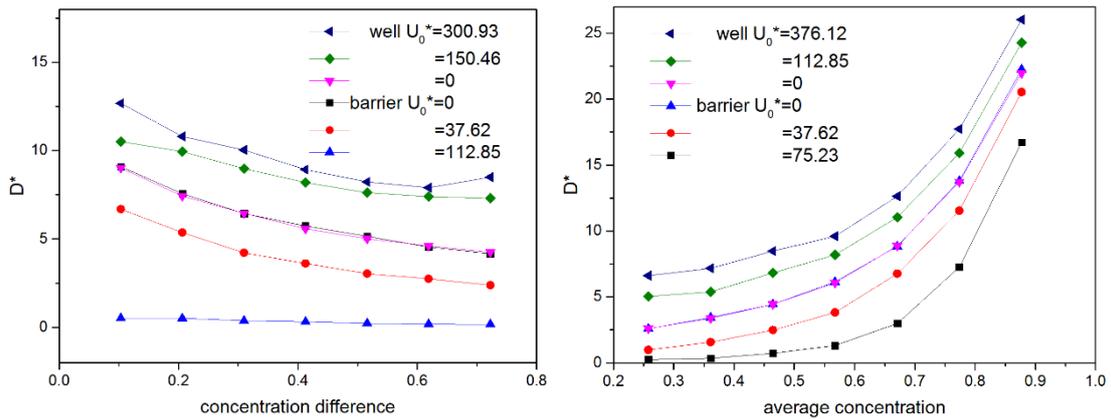

Fig 8: In the graph of the concentration difference, even though all the lines are decreased, the data of flux also show that particle flux is increased with the enlargement of concentration difference; stabilizing more particles, the programs for the right graph are set the number of time steps as $5*10^9$.

These two pictures about concentration change elaborate how diffusion relies on concentration difference and average concentration, under the list of several potentials' manifestation. From the left picture about the effect of the concentration difference, the diffusion coefficient is decreased with the enlargement of concentration difference. The D tendency is a curve approaching to the horizontal, and the more upper curve is more bent. In the right side, the diffusion coefficient is increased when increase average concentration. Both two graphs present barriers and wells have the same tendency, but the diffusion ability of the potential well is better than that of barriers.

The graphs about the concentration can be explained by the entropy in thermodynamics. In the brilliant relationship of Boltzmann, $S = k \ln \Omega$, in which S is the entropy, $\Omega$ is the number of state. It indicates the diffusion coefficient is increased with average concentration. In the case of concentration difference, the Fick's law tells us the difference in concentration is the favorable condition for transport, as the date of flux show.

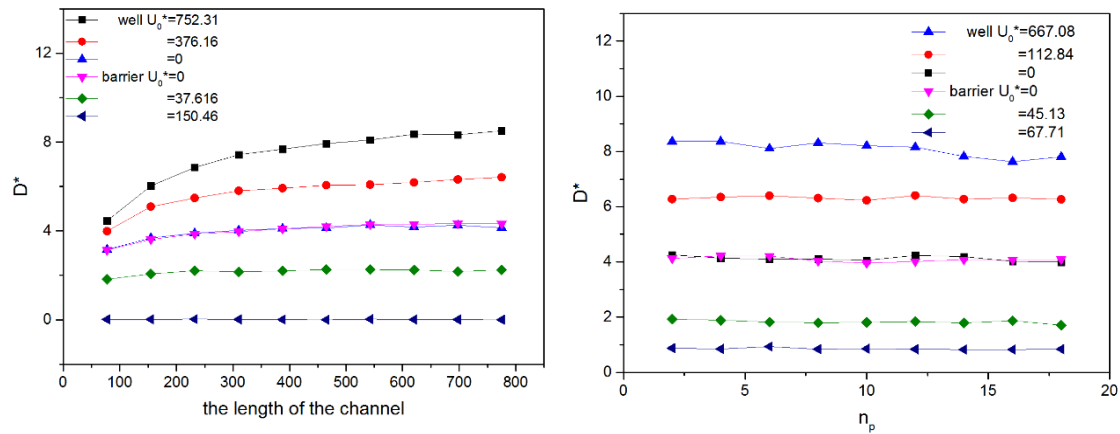

Fig 9: these two pictures imply neither the length of a transport channel nor the period of the force field can affect the transport obviously. For this part, the potential field need to be amended for the sake of invariant of the potential depth or height though change variables.

From the left graph there is a small curl in the area of relatively short length channel which implies ballistic transport, but it has a steady value in long channel. In the period graph and the most part of the length graph, the diffusion coefficient goes steadily.

Some linear fittings have been done and exhibit the channel length and the force period really work in the collective diffusion, but it is the other matter about the dynamics, not the statistics characteristics in the relatively long channel.

# Conclusion

Finally, it can comes to the conclusion about the collective transport under the external potential. Our work builds a microcosmic model to simulate particles diffusion process with the help of some usual algorithms. The diffusion coefficient would be figured out and then the pictures about that could be draw up under the control of some series of condition. On all of these diagrams, there is one common – the situations of wells and barriers coincide together when the external force becomes zero, testifying the validity of the results. The phenomena from the results can be explained by canonical ensemble distribution, Boltzmann relationship, Einstein relation, etc.

Potentials really affect the transport even though additional potentials do not change the potential energy of the system at all. Two contrary potentials – the potential well and the barrier – represent two contrary conditions to the transport. All of these diagrams indicate that the potential well is an active drive for particles collective diffusion and the potential barrier is an obstacle inversely.

The system is combined with different fields, which external force field, concentration field and particle potential couple and tangle together. Also, statistics analysis and subsection method are used to divide flux into density and velocity. External potential impacts the distribution function of Gibbs relation, but L-J potential impede in some extent. Averagely to say, the higher particle density means the larger particle transport.

In the participation, general acknowledged principle, Maxwell velocity distribution, still applies in the approximate equilibrium state.

Other surroundings such as the temperature, the concentration difference, the average concentration are the subordinating parameters, so whatever the type of the potential, the coefficients have the same tendency. Considering the length of the transport channel and the period of the external field, they have the delicate influence on diffusion.

The academic significance of this research is enriching the nonequilibrium theory, verifying the relevant knowledge and getting theoretical explains. In the subcellular level, particles always confront with the external force and move. This transport model also base on the simulation, and we hope these works can give some references about numerical needs in the homologous issue like controlling the flux by controlling the external potential. Particularly in biology, for instance, the selectivity filter or protein channel are such that physical systems. Creatures live in the nonequilibrium state. Fick's law, as the popular transporting law, still faces the difficult situation to solve any real nonequilibrium problem especially in biology. With the help of that diffusion simulation, it may be a guide or reference for the control of molecular transport.

# Acknowledgements

The authors would like to thank the high performance computing platform in Jinan University and Siyuan clusters. Thanks to Yong-jian zeng for some graphs and calculation. This work was supported in part by the Natural Science Foundation of Guangdong Province, China (Grant no. 2014A030313367); and the Fundamental Research Funds for the Central Universities, JNU (Grant no. 11614341).